\documentclass[article]{IEEEtran}
\usepackage{pgfplots}
\usepackage{pgfplotstable}
\usepackage{multicol} 
\pgfplotsset{compat=1.17}
\usepackage{tikz}
\usepackage{ifpdf}
\usepackage{amsthm}

\usepackage{cite}

\usepackage{subcaption}

\makeatletter	
\def\@IEEEpubidpullup{8\baselineskip}	
\makeatother	

\begin{document}

\title{TraderTalk: An LLM Behavioural ABM applied to Simulating Human Bilateral Trading Interactions}

\author{\IEEEauthorblockN{Alicia Vidler}
\IEEEauthorblockA{UNSW,
Email: a.vidler@unsw.edu.au}
\and
\IEEEauthorblockN{Toby Walsh}
\IEEEauthorblockA{UNSW,
Email: t.walsh@unsw.edu.au}}

\maketitle

\IEEEoverridecommandlockouts	\IEEEpubid{	\parbox{\columnwidth}{\vspace{-2\baselineskip} 
~\copyright 2024 IEEE. Personal use of this material is permitted.  Permission from IEEE must be obtained for all other uses, in any current or future media, including reprinting/republishing this material for advertising or promotional purposes, creating new collective works, for resale or redistribution to servers or lists, or reuse of any copyrighted component of this work in other works. October 10th, 2024 \hfill}	\hspace{0.3\columnsep}\makebox[\columnwidth]{\hfill}}	\IEEEpubidadjcol	

\begin{abstract}
We introduce a novel hybrid approach that augments Agent-Based Models (ABMs) with behaviours generated by Large Language Models (LLMs) to simulate human trading interactions. We call our model TraderTalk. Leveraging LLMs trained on extensive human-authored text, we capture detailed and nuanced representations of bilateral conversations in financial trading. Applying this Generative Agent-Based Model (GABM) to government bond markets, we replicate trading decisions between two stylised virtual humans. Our method addresses both structural challenges—such as coordinating turn-taking between realistic LLM-based agents—and design challenges, including the interpretation of LLM outputs by the agent model. By exploring prompt design opportunistically rather than systematically, we enhance the realism of agent interactions without exhaustive overfitting or model reliance. Our approach successfully replicates trade-to-order volume ratios observed in related asset markets, demonstrating the potential of LLM-augmented ABMs in financial simulations.
\end{abstract}

\IEEEpeerreviewmaketitle

\section{Introduction}
Large Language Models (LLMs) have garnered much attention since 2022, and continue to evolve rapidly, demonstrating capabilities in understanding and generating human-like text across various domains.  Integrating LLMs into multi-agent frameworks is seen as a key future design method for AI  \cite{NgA}, aiming to replicate complex human interactions and decision-making processes \cite{park2023generative}. Behaviours like risk inertia, aversion, and ambiguity avoidance significantly impact financial markets \cite{NBERw29915}, and can be so extreme as to dissuade traders from transacting all together \cite{ambiguityBossaerts}. In this paper we investigate whether an LLM, managed by an agent, can simulate human behaviours in bilateral asset trading. 

ABMs can effectively simulate interdependent, adaptive complex systems \cite{Bai_2020}, \cite{Gilbert_2007}, however, their design involves significant parameterisation of agent behaviours. Logic-based methods like "belief, desire, intention" \cite{Wooldridge_2009}, \cite{Kirilenko_2017}, \cite{Paulin_2019} are common, though agent feature specification, especially logic and decision-making is challenging \cite{Fehler_2006} and no consensus exists on calibration methods \cite{avegliano2019using}.


ABMs are well-established in financial market simulations, with the necessity of heterogeneous agents recognised \cite{Hayes_2014, Paddrik_2012}. The rapid evolution of LLMs—such as OpenAI's GPT-4o and GPT-o1 in 2024—poses challenges for systematic testing, as updates can render studies obsolete. To address this, we propose a flexible framework using the most current, widely accessible LLM (GPT-40-mini) without relying on specific model versions or fine-tuned models. Due to rapid advances and inherent lack of transparency in models like ChatGPT, we present limited results as a proof of concept.

We apply our methods to bilateral trading in government bond markets, such as UK Gilt bonds.  This market involves participants such as market makers (MMs), clients, and inter-dealer brokers \cite{BOE}.  Systemically important to the countries they serve, in markets such as Australia and the UK, MMs facilitate most government bond transactions, which occur over-the-counter (OTC) with limited publicly available data \cite{Cheshire2015, Pinter2023}. Thus, modelling of these markets requires novel methods of ABM design and enhancements \cite{vidler2024modellingopaquebilateralmarket}. By incorporating LLMs and focusing on negotiation and decision-making, our framework offers new opportunities to simulate realistic human interactions in these markets, aiming to enhance methodologies and provide more nuanced, realistic market simulations.

We introduce TraderTalk, a bespoke generative agent-based model (GABM) that integrates a general-purpose LLM into ABMs using open-source software, Concordia \cite{vezhnevets2023generative} and LLM prompting methods. By injecting human-like behaviours and uncertainties into logic-based ABMs without domain-specific tuning, we aim to enhance simulation realism in bilateral financial trading. This paper is structured as follows: we discuss recent research on LLMs and ABMs and related concerns, present our architecture for integrating LLMs in ABMs, show test case results for financial trading scenarios, and conclude with future research directions.

\section{Recent Research and Relevant concerns}
The potential of LLMs is well acknowledged, however limitations on numerical reasoning, and prompting methods persist \cite{srivastava2023imitationgamequantifyingextrapolating, zhang2024llmmastermindsurveystrategic}, in particular mathematical reasoning \cite{ahn-etal-2024-large}.  Prompt design remains an active research area with many challenges and opportunities; even simple prompts to "re-read" input are found to significantly improves performance \cite{xu2024rereadingimprovesreasoninglarge}.  Complicating things further is the finding that non-AI experts often adopt "opportunistic rather than systematic approaches" to prompt design \cite{Zamfirescu-Pereira2023}.  Thought-eliciting methods, like Chain-of-Thought (CoT), are popular as they aim to "elicit the reasoning process in the output" \cite{ChainOfThoughtWei2022}, relying on giving an LLM "worked examples", drawing inspiration from human learning theories.  We build upon these concepts, incorporating an agent into this bi-directional process and COT, within a simulated conversation. 

\subsection{Application and Financial asset trading}
Current financial market simulations using LLMs prioritise price dynamics over trading activity \cite{Xu2023}. In bond markets, prices are largely known due to interest rate assumptions and are heavily influenced by monetary policy \cite{Fabozzi}; moreover, MMs are legally required to maintain a minimum market share of government bonds \cite{BOEOpps}. Thus, liquidity---the movement of assets between parties---is a key concern.  Modelling these asset flows and transaction intentions remains an active, though limited, research area \cite{vidler2024modellingopaquebilateralmarket}. LLMs could be particularly useful in markets with limited data and dominant bilateral trading interactions. Despite their potential, the application of LLMs in simulating human-to-human interactions in financial markets, especially for bilateral trading, remains under-explored. Our research addresses this gap.



\subsection{Generative Agent Based Models}
Using Concordia \cite{vezhnevets2023generative} we integrate LLMs with ABMs to create Generative Agent-Based Models (GABMs), enabling agents to "apply common sense" and "act reasonably" within simulated environments \cite{park2023generative}. A key feature is the Game Master agent, which translates natural language requests into executable actions. However, recent studies \cite{Abdelnabi, zhang2024llmmastermindsurveystrategic} reveal that current LLMs under perform in negotiation tasks and strategic reasoning within agent-based systems. To address these challenges, we focus on enhancing agents' negotiation and decision-making capabilities by integrating LLMs beyond traditional rule-based ABMs \cite{Wooldridge_2009}. Unlike prior work \cite{WuAutoGen} that designs generic agents using LLMs, we concentrate on design features essential for decision-making through negotiation. By augmenting agents with LLMs, we aim to create more flexible and adaptable decision-making processes in complex environments. Our work demonstrates that even in simple settings, combining LLMs with agents can enhance realism in ABM models, benefiting future research.



\subsection{Order to Trade Ratio (OTR): Uncertainty in Human-Directed Trading}
Financial trading involves significant uncertainty, often requiring multiple attempts before a trade is executed—even in transparent equity markets. In 2024, the average OTR for major U.S. equity exchanges was approximately 4.61\% \cite{SEC_24ExchangeTradeOrderVolume}. Thus, up to 96\% of daily trading requests do not result in trades, complicating the modelling of human behaviour \cite{SEC_24ExchangeTradeOrderVolume}. Work by \cite{farmer2012minimum} explore possible theories and impacts, including market spoofing \cite{DalkoOTRreg}, and theories of ambiguity aversion in human trading are discussed in \cite{ambiguityBossaerts} and \cite{NBERw29915}. The causes of high OTR's remain an open research question, though we aim to leverage LLMs to include this aspect for added realism.

\section{\textbf{TraderTalk}: Architecture and Results}

\textit{TraderTalk} is an ABM featuring two market making agents, "Josephine" and "David," each with initial characteristics such as bond holdings and explicit trading intentions. Our model passes information between the agents and an external LLM, consistently using GPT 4o-mini throughout our experiments. We address two research questions:

\subsection*{\textbf{Research Question 1: Can an LLM realistically and appropriately respond in a bilateral trading interaction? (RQ1: Baseline)}}

We initially implement our model with agents functioning as messengers passing basic information to the LLM. The simulated scenario involves the first MM initiating contact with another MM, who does not wish to trade because they are not a buyer. Using a CoT framework, agents are guided through a sequence where they:

\begin{itemize}
    \item Summarise new information.
    \item Clarify their roles and objectives.
    \item Assess their current bond holdings.
    \item Decide whether to trade or not.
\end{itemize}

If a trade is decided, they determine the appropriate action to meet their obligations, such as buying or selling bonds, flattening their trading book, or maintaining their current position. The agents are initialised with prompts derived from the CoT, which in turn drives a simulated conversation, each agent responding (and concluding) based on LLM reasoning. If they choose to trade, the LLM is asked to select from 4 possible options (buying or selling bonds, flattening their trading book, or no trade). After both agents have contributed and made a selection, the conversation text is separately analysed to determine if a trading decision was reached.  In this setup, the ABM provides only a premises to the LLM, which independently makes decisions. 
 
Our goal is to evaluate how often the LLM correctly reasons to produce a simulated conversation resulting in "no trade". We selected the "no trade" decision to avoid complexities related to numerical reasoning \cite{ahn-etal-2024-large}, focusing on a scenario where a trader holding no bonds (a flat position) is expected to follow the straightforward, implicit prompt of not trading. We isolate the LLM's ability to interpret and apply a specified trading intention. The CoT prompt is thus structured so that the correct outcome is for the agent to choose "no trade" from the final multiple-choice options provided it.

\begin{figure}  
    \centering
    \begin{subfigure}{.35\columnwidth}  
        \includegraphics[width=\linewidth]{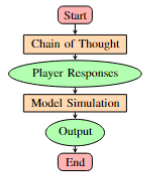}
        \caption{RQ1 Baseline}
        \label{fig:COT_traditional}
    \end{subfigure}%
    \begin{subfigure}{.54\columnwidth}  
        \includegraphics[width=\linewidth]{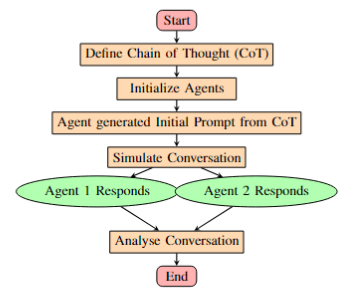}
        \caption{RQ2: GABM with Concordia acting as agent handler}
        \label{diagram_GABMRQ2}
    \end{subfigure}
    \caption{Model Architectures}
        \label{figure}
\end{figure}

\textbf{Trading Premise} \textit{"You are a market maker for UK gilts responsible for providing liquidity in the UK government bonds.  You are supposed to at all times hold 0 bonds. Today, you actually have 0 bonds, which means your holding is actually flat" }.

\subsection*{Results and Conclusion}
Across 300 simulations, the LLM correctly decided to select not to trade in 180 instances, demonstrating that the LLM can reason and follow the agent-based intention 60\% of the time.  We observe no significant differences with smaller sample sizes.  This test isolates the LLM's ability to reason given a natural language trading intention. There is an absence of any direct comparison of how frequently human traders perfectly follow such intentions however.  Furthermore, existing research on human decision-making highlights ambiguity and the distinction between following rules and intentions, suggesting that achieving a 100\% success rate is unrealistic, although this is not quantifiable.  We use this result as a baseline for future tests in more complex scenarios and believe that this achieves the goal of including human like attributes, though the quantum of such is beyond the scope of this work.

In the remaining 40\% of simulations where the LLM did not follow the correct intention, 23.6\% involved the LLM attempting to "flatten" a trading book that was already flat, suggesting a misunderstanding or failure to adhere to the instructions and market parlance. The other 16.4\% of responses reflected active trading positions that directly contradicted the premise, with the LLM expressing a desire to "buy" in 10\% of cases and to "sell" in 6.3\% of cases. We interpret this variation in behaviour introduced by the LLM as analogous to the emergence of unexpected properties within traditional agent-based models.

\subsection*{\textbf{Research Question 2 (RQ2): TraderTalk—Can a GABM Make a Trading Decision in a Realistic Manner?}}

In this test, we enhance the framework from RQ1 by passing specific agent information to the LLM using Concordia's agent-handling mechanism \cite{vezhnevets2023generative}. Each agent is initialised with distinct roles: this time David holds a negative bond position and needs to buy bonds, while Josephine holds a positive bond position and needs to sell bonds. Concordia's Game Master design facilitates these interactions, functioning as a meta-agent manager that supervises exchanges and ensures smooth decision-making. Consequently, the model design is augmented from Figure \ref{fig:COT_traditional} to produce Figure \ref{diagram_GABMRQ2}.

The new process is as follows:

\begin{enumerate} 

\item \textbf{Define Chain of Thought (CoT):} Use \textbf{Project Context}

\item \textbf{Initialise Agents:} Assign specific roles and initial conditions to the agents (see \textbf{Trading Roles} below).

\item \textbf{Generate Initial Prompts:} Agents, via the Game Master, generate their own responses to the CoT questions from RQ1, stored and passed to the LLM.

\item \textbf{Simulate Conversation:} Managed by the Game Master, the LLM simulates the dialogue between the agents, with each responding in turn based on previous interactions and their trading objectives.  The conversation continues until the Game Master determines it has concluded

\item \textbf{Analyse and Conclude:} .We then analyse the conversation for trade occurrences, quantities, and dialogue content.

\end{enumerate}

\textbf{Project Context:}
    \textit{"You are a market maker for UK gilts responsible for providing liquidity in the UK government bond. Your job is to answer incoming queries from other market makers to buy and sell UK government bonds by considering if you wish to do so. UK government bonds trade at mid price. You aim to make a trading decision in every conversation, either buy , sell or decline to trade. You must act professionally in your conversations, and any decision you take is clearly communicated to the other party and you repeat what is agreed."}

\textbf{Trading Roles:}

\textbf{David}: \textit{"You are a market maker for UK gilts responsible for providing liquidity in the UK government bond, you are supposed to at all times hold 0 bonds.  Today, you actually have negative 10 million worth of bonds, your role is to buy the bonds if you have a negative holding"}

\textbf{Josephine}: \textit{"You are a market maker for UK gilts responsible for providing liquidity in the UK government bond, you are supposed to at all times hold 0 bonds. Today you have 10 million worth of bonds, your role is to sell bonds if you are a holder, you need to call another market maker to trade away your bonds"}

Unlike in RQ1, where the LLM operated independently, this setup integrates the ABM into decision-making to evaluate how often the LLM produces simulated conversations with correct reasoning regarding trade intentions and executions. Agents directly inform action direction (identifying buyers and sellers).

\subsection*{Results and Conclusion}
Again, we conducted 300 simulations using GPT 4o-mini; we see agents intended to trade in \(58\%\) of cases, and at least one party was willing to trade in \(98\%\) of instances. Agent "Josephine" closely aligned with her role, intending to trade \(97.3\%\) of the time, while "David"'s intention was lower at \(58.7\%\); he explicitly declined to trade in \(22.3\%\) of responses, and \(19\%\) were unclear.  The \(58\%\) rate at which both parties intended to trade is close to the \(60\%\) correct response rate in our RQ1, suggesting consistent reasoning abilities of the LLM across different model designs in RQ1 and RQ2.

Despite the high intention to trade, actual trades occurred in only \(5.7\%\) of cases, highlighting a significant gap between intentions and execution.  While LLM-driven agents often desire to trade, the necessary LLM dialogue needed to finalise a trade seems less frequently generated, producing low successful trading rates - align with real-world observed  OTR levels \cite{SEC_24ExchangeTradeOrderVolume}. \textit{TraderTalk} only identified the correct initial bond holdings for both parties in 2.34\%, with 32\% of responses omitting starting values altogether, reflecting difficulties in recalling initial numerical conditions, in line with \cite{ahn-etal-2024-large}.  Overall, our ABM augmented with LLM behaviours (\textit{TraderTalk}) appears capable of producing interactions consistent with sparse real-world data and making trading decisions in a realistic manner.

\section{Conclusion}
We present \textit{TraderTalk}, a novel LLM behavioural agent-based model that simulates realistic human bilateral trading interactions without extensive model tuning. Utilising a state-of-the-art, non-domain-specific, non-fine-tuned LLM within the Concordia framework (GTP 4o-mini), we demonstrate limited yet realistic trade negotiations (\textbf{RQ1}), interpretation, and trade execution decisions (\textbf{RQ2}) at frequencies approximating those in U.S. equity markets. By addressing key challenges like coordinating agent turn-taking, our simulation achieves a trade-to-order ratio similar to real markets. Discrepancies between trading intentions and execution in agent outputs enhance the realism of stylised human traders, capturing decision-making processes in bilateral trading environments where much interaction occurs outside formal exchanges. This proof-of-concept indicates that LLMs can meaningfully enhance the realism of behavioural simulations in ABMs for financial market modelling and provides a foundation for future research into more complex multi-agent and multi-market simulations.  Future work should enhance GABM's understanding of implicit trading rules and dynamic market conditions; by refining their ability to capture human decision-making, LLMs could offer more robust simulations for policymakers, regulators, and market participants alike.



\section*{Acknowledgement}
We thank Dr Arnau Quera-Bofarull and Dr Nick Bishop (Oxford University), for their guidance on the research topic. We thank an anonymous market maker for their input.  This work is funded in part by an ARC Laureate grant FL200100204.

\bibliographystyle{IEEEtran}
\bibliography{Main_mybibliography}

\end{document}